\documentclass[10pt,letterpaper]{article}
\usepackage[top=0.85in,left=2.75in,footskip=0.75in]{geometry}
\usepackage{ulem}
\renewcommand{\sout}[1]{\unskip}

\usepackage{amsmath,amssymb}
\usepackage[finalnew]{trackchanges}

\usepackage{changepage}

\usepackage[utf8x]{inputenc}

\usepackage{textcomp,marvosym}

\usepackage{cite}

\usepackage{nameref,hyperref}

\usepackage[right]{lineno}

\usepackage{microtype}
\DisableLigatures[f]{encoding = *, family = * }

\usepackage[table]{xcolor}

\usepackage{array}

\newcolumntype{+}{!{\vrule width 2pt}}

\newlength\savedwidth

\raggedright
\usepackage[aboveskip=1pt,labelfont=bf,labelsep=period,justification=raggedright,singlelinecheck=off]{caption}

\makeatletter
\renewcommand{\@biblabel}[1]{\quad#1.}
\makeatother

\usepackage{lastpage,fancyhdr,graphicx}
\usepackage{epstopdf}
\fancyheadoffset[L]{2.25in}
\fancyfootoffset[L]{2.25in}
\newcommand{\dd}{\mathrm{d}}
\graphicspath{{./figs/}}

\begin{document}
\vspace*{0.2in}

\begin{flushleft}
{\Large
\textbf\newline{Regularized Bayesian calibration and scoring of the WD-FAB IRT model improves predictive performance over  marginal maximum likelihood} 
}
\newline
\\
Joshua C. Chang  \textsuperscript{1},
Julia Porcino\textsuperscript{1},
Elizabeth K. Rasch\textsuperscript{1},
Larry Tang\textsuperscript{1,2*} \\
\bigskip
\textbf{1} Rehabilitation Medicine Department, NIH Clinical Center, Bethesda, \remove{MD}\add{Maryland, United States of America}
\\
\textbf{2} National Center for Forensic Science, Department of Statistics, University of Central Florida, Orlando, \remove{FL}  \add{Florida, United States of America}
\bigskip

%
%

  
 \add{* Corresponding author}
 
 \add{*E-mail:} tangl5@cc.nih.gov

\end{flushleft}
\section*{Abstract}
Item response theory (IRT) is the statistical paradigm underlying a dominant family of generative probabilistic models for test responses,
used to quantify traits in individuals relative to target populations.
The graded response model (GRM) is a particular IRT model that is used for ordered polytomous test responses.
Both the development and the application of the GRM and other IRT models require statistical decisions.
For formulating these models (calibration), one needs to decide on methodologies for item selection, inference, and regularization.
For applying these models (test scoring), one needs to make similar decisions,
often prioritizing computational tractability and/or interpretability.
In many applications, such as in the Work Disability Functional Assessment Battery (WD-FAB),
tractability implies approximating an individual's score distribution using estimates of mean and variance,
and obtaining that score conditional on only point estimates of the calibrated model.
In this manuscript, we evaluate the calibration and scoring of models under this common use-case using Bayesian cross-validation.
Applied to the WD-FAB responses collected for the National Institutes of Health, 
we assess the predictive power of implementations of the GRM based on their ability to yield, on validation sets of respondents, \sout{estimates of latent ability with uncertainty} \add{ability estimates} that are most predictive of patterns of item responses.
Our main finding  indicates that regularized Bayesian calibration of the GRM outperforms the \sout{prior}\add{regularization}-free empirical Bayesian procedure of marginal maximum likelihood. We also motivate the use of compactly supported priors in test scoring.


\section*{Introduction}
Item response theory (IRT) encompasses a class of latent variable models for quantifying traits,
such as abilities and attitudes, using questionnaires~\cite{changItemResponseTheory2005,friesItemResponseTheory2014}. 
Some of its highest-profile applications include the Graduate Record Exam (GRE)~\cite{kingstonFeasibilityUsingItem1982} and the Scholastic Aptitude Test (SAT)~\cite{carlsonItemResponseTheory2013}. 
 Besides its widespread use in high-stakes educational assessment,
 it is also heavily used in psychometrics~\cite{altemeyerEnemiesFreedomUnderstanding1988,goldbergDevelopmentMarkersBigFive1992} and medical diagnostics~\cite{sullivanAssessmentAlcoholWithdrawal1989}.


Fundamentally, IRT models are generative non-linear factor analysis models~\cite{changProbabilisticallyautoencodedHorseshoedisentangledMultidomain2019}.
 These models yield predictions, in the form of probability mass functions, for how a particular person will respond to a particular test item.
 The key assumption in IRT is that the probability of one's response to any particular item on a test is a function composed of person and item-specific effects.
 As commonly implemented, the item-specific parameters inform each item's difficulty and discriminatory power.
 The person-specific parameter relates to  an underlying ability relative to a target population.
 Altogether, these models assume that an individual can be characterized by a low-dimensional set of parameters known as traits or abilities, \add{which are the embedded factors}. Hence, these parameters constitute a \emph{representation} of an individual's traits, or interchangeably, a person's responses.

\subsection*{Statistical choices in IRT model construction}

IRT models are informed through a process known as \emph{item calibration}.
In calibration, one aims to train the item parameters in the IRT model using responses from a sample of the target population.
In order to do so, however, one must \sout{also } \add{either} simultaneously infer \add{or control for} the person-specific parameters within the sample.
Hence, calibration involves the parallel inference tasks of determining person and item-specific parameters.

\emph{Scoring} is the process by which calibrated IRT models are applied to new responses,
in order to obtain person-specific parameters applicable to the new respondents.
In scoring, the new respondents are fit into the scale defined by the original calibration responses.
Hence, the application of IRT models for use in prediction differs from usual statistical
models in that another inference step is required after the initial training.

As high-dimensional models, the details of how one calibrates and scores IRT models are important.
Unregularized maximum likelihood is known to be insufficient for calibration, due to parameter unidentifiability confounded by the nonlinear nature of  the models. 
Similarly, scoring is sometimes done in settings where few test responses are known and maximum likelihood
is unstable -- this occurs regularly in computer adaptive testing for instance, where a test
is scored in real-time, and this scoring informs the presentation of new items. 

This particular drawback of the maximum likelihood method can be found in    the Work Disability Functional Assessment Battery (WD-FAB) which is a relatively recent application of IRT to the assessment of
work-related \add{physical and behavioral} function intended to inform processes related to disability determination and other potential applications \cite{marinoWorkrelatedMeasuresPhysical2015}.

\subsection*{Work Disability Functional Assessment Battery (WD-FAB)}

We focus on the concrete application of improving the statistical properties of the IRT model underlying the Work Disability Functional Assessment Battery (WD-FAB)~\cite{marinoWorkrelatedMeasuresPhysical2015,jetteWorkDisabilityFunctional2019,marfeoDevelopmentInstrumentMeasure2013,marfeoDevelopmentNewInstrument2016,marfeoImprovingAssessmentWork2018}.

\subsubsection*{Background}

 The concept of work disability is evolving as represented by the World Health Organization’s International Classification of Functioning, Disability and Health (World Health Organization [WHO], 2001). Modern models of work disability characterize the outcome of the interaction  of a person’s functional abilities within the work environment.  Due to the complex nature of the interaction, a fundamental issue is how to identify and measure work disability within this contemporary framework ~\cite{marinoWorkrelatedMeasuresPhysical2015}.  

The U.S. Social Security Administration (SSA) provides support to adults and children who qualify as disabled through the Supplemental Security Income (SSI) and Social Security Disability Insurance (SSDI) programs that provide health insurance and cash benefits to beneficiaries.  As these are the two largest federal disability programs in the U.S. supporting millions of Americans, accurate  assessment of work disability is critical to applicants as an important safety net program and to  the federal government to effectively allocate resources.  The Social Security Administration (SSA) uses a statutory definition of work disability characterized as the inability to take part in ``substantial gainful activity due to any medically determinable physical or mental impairment that can be expected to result in death or to last for a continuous period of not less than 12 months.''  The Work-Disability Functional Assessment Battery (WD-FAB) was developed as an additional, comprehensive source of information about whole person function intended to support adjudicators when making disability determinations and/or re-determinations.

\subsubsection*{Prior development of the instrument}

In this paper we evaluated the Work Disability Functional Assessment Battery (WD-FAB)  which was developed by researchers at the Boston University Health and Disability Research Institute (BU) in conjunction with the National Institutes of Health (NIH). It is a computer-adaptive testing tool, backed by an IRT model, encompassing eight scales including four physical function scales and four mental function scales to identify self-reported function relative to work. 
 The items within these scales consist of Likert-scaled multiple choice questions.

Work-related job function is a multifaceted concept, not easily summarized by any single quantitative factor. In the development of the WD-FAB, a combination of expert guidance and empirical evidence was used to inform the multidimensional nature of the instrument. An expert panel developed an item bank consisting of Likert scaled questions relating to physical and mental function. In a series of three surveys, a large-scale simple random sample of approximately 5000 SSA disability claimants in the United States was administered the entire item bank for use in calibrating the instrument. Separately, approximately 2000 individuals from the general population of both claimants and non-claimants were   in placing the instrument in larger population context (in other words, a normative or reference sample).

The empirical portion of the creation of the WD-FAB, from individual responses, proceeded in two steps.
The first step, exploratory factor analysis~\cite{fergusonExploratoryFactorAnalysis1993,goretzkoExploratoryFactorAnalysis2019, williamsExploratoryFactorAnalysis2010,howardReviewExploratoryFactor2016} (EFA) splits up an item bank into
independent domains. The second step, confirmatory factor analysis~\cite{vanprooijenConfirmatoryAnalysisExploratively2001} (CFA), verifies that a latent
unidimensional trait is explanatory for the pattern of responses found for questions in each
domain. 

 In EFA, step-wise item selection based on p-value cutoffs were used to factor the items into eight domains.  Four of the domains pertain to physical function: Basic Mobility (BM), Fine motor function (FMF), Upper body function (UBF), and Community Mobility (CM). 
 Four of the domains are used to evaluate psychological characteristics: Mood and emotions (ME), Resilience (RS), Self Regulation (SR), and Communication and Cognition (CC).
 
 In CFA, heuristic (and ultimately arbitrary) values for fit statistics based on the PROMIS guidelines~\cite{friesItemResponseTheory2014,dewaltEvaluationItemCandidates2007} justified the instrument. In neither the EFA nor the CFA step were generalizability of the instrument considered. Additionally, both EFA and CFA, based on their own factorization models, do not guarantee consistency with nonlinear IRT models, which are themselves factor models.
 The specific IRT model adapted for the WD-FAB is the graded response model~\cite{samejimaEstimationLatentAbility1969} (GRM).
 With the domains in place, marginal maximum  likelihood (MML) was used to calibrate eight independent IRT models, and Warm's weighted maximum likelihood~\cite{warmWeightedLikelihoodEstimation1989} was adopted for test scoring.

\subsection*{\sout{Our contributions}\add{Purpose of the present study}}

In this manuscript we improve on the predictive power of the graded response model \add{underlying the WD-FAB} by regularizing its calibration  \add{using fully-Bayesian methods. This approach is motivated by prior literature showing Bayesian calibration to be more accurate than MML calibration for small datasets, even while using diffuse priors}~\cite{kieftenbeldRecoveryGradedResponse2012,lordMaximumLikelihoodBayesian1986}, \add{ and the modern trend in applied Bayesian statistics towards utilizing stronger weakly informative priors for the purpose of statistical regularization}~\cite{gelmanPriorCanOften2017,shengInvestigatingWeaklyInformative2017,lemoineMovingNoninformativePriors2019}.

As a metric \add{for assessing predictive power}, we adapt cross validation to estimate out-of-sample \sout{log likelihoods} \add{model accuracy} in a manner that is consistent with how the WD-FAB instrument is used and interpreted. \add{For predictive model assessment, cross validation and related methods} are commonly used in machine learning and have found widespread adaption
in the Bayesian statistical modeling world, but are not in common use for IRT model assessment. 

Commonly-used \sout{procedures for doing so} \add{ alternatives to cross validation include information criteria such as the }  WAIC~\cite{watanabeAsymptoticEquivalenceBayes2010,watanabeWidelyApplicableBayesian2013,luoPerformancesLOOWAIC2017} \add{and the AIC, which under certain conditions}~\cite{gelmanUnderstandingPredictiveInformation2014,stoneAsymptoticEquivalenceChoice1977} \add{ are asymptotically equivalent to leave-one-out cross validation.}
\add{However, it is not straightforward to compare Bayesian versus non-Bayesian models using standard information criteria.  Additionally, information criterion require two models to have identical prediction outputs for direct comparison and are therefore not flexible enough to handle common IRT model selection use cases like item selection.  Finally, these criteria are not typically mindful of how models are interpreted.}

To be mindful of common IRT-model use cases, \sout{where ability estimates are parameterized by mean and variance, and where only point estimates of item parameters are used in scoring, we develop a custom variant of the out-of-sample leave-K-out log likelihood.} \add{where both model calibration and scoring require statistical decisions, and scores are interpreted alongside their inferred errors, we develop a custom variant of a cross validation metric based on the out-of-sample leave-K-out log marginal likelihood.}
We show that Bayesian IRT model calibration, coupled with regularized Bayesian scoring, outperforms the commonly-used procedures of marginal maximum  likelihood (MML) calibration and weighted likelihood estimation (WLE) scoring.

\section*{Materials and methods}
\begin{figure}[!h]
    \centering
    \includegraphics[width=10pc]{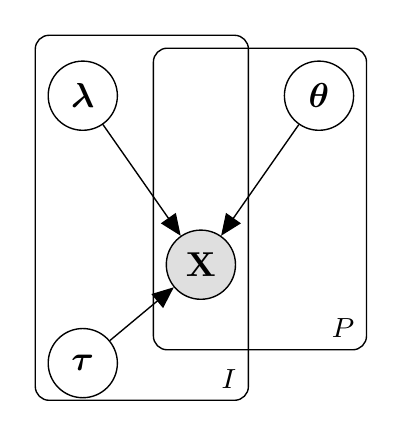}
    \caption{{\bf Graded response model.} 
        The  Graded response model  (GRM) with item parameters
        $\boldsymbol{\lambda}, \boldsymbol{\tau},$ and person-specific parameters
        $\boldsymbol{\theta}.$}
    \label{fig:grm_plate}
\end{figure}

Although the methods in this paper generalize broadly to other item response theory models,
we formulate  our methods based on the unidimensional graded response model (GRM) for polytomous item responses.
The GRM~\cite{samejimaEstimationLatentAbility1969} applies to assessments where there is an intrinsic ordering in responses, as in Likert scales.
According to the GRM,
the probability of a response of $j$ to item $i$ for person $p$ obeys the likelihood function
\begin{eqnarray}
\Pr( X_{pi} = j \vert \theta_p, \boldsymbol\tau , \boldsymbol\lambda  )  &= \Pr(X_{pi}\geq j \vert\theta_p, \boldsymbol\tau , \boldsymbol\lambda )- \Pr(X_{pi}\geq j+1 \vert\theta_p, \boldsymbol\tau , \boldsymbol\lambda ) \nonumber\\
&= \frac{1}{1 + \exp(\lambda_i( \tau_{ij}-\theta_p )) } - \frac{1}{1 + \exp(\lambda_i(  \tau_{i,j+1} -\theta_p )) } 
,\label{eq:GRM}
\end{eqnarray}
where $\lambda_i$ are the item-specific discrimination parameters and $\tau_{ij} < \tau_{i,j+1}$ are the item-specific 
threshold parameters, and $\theta_p$ are person-specific ability parameters. 
The schematic of the GRM is presented in Fig \ref{fig:grm_plate}, where for each of $P$ people, predictions of their $I$ item responses are determined by these parameters.
 
 Eq.~\ref{eq:GRM}, demonstrates that in the GRM, the person-specific parameter $\theta_p$ only has interpretability relative to the item-specific parameters. 
 Mathematically speaking, the likelihood function in Eq.~\ref{eq:GRM} is unidentifiable, in both location and scale. 
 To improve identifiability, one typically either imposes or encourages a scale and location on the person-specific $\theta_p$,
 on the population level, so that a target population has $\theta_p$ follows roughly a Gaussian distribution of a given scale (typically unit).

The unidimensional formulation of Eq.~\ref{eq:GRM} can be extended to multidimensional traits
 (representing each individual using a vector of values rather than a scalar),
by fitting models for each domain (dimension) separately, effectively partitioning
tests into domains of items \add{(as per the PROMIS guidelines}~\cite{PromisePROMIS2010}). Such  item partitioning is typically completed with the aid 
of linear factor analysis methods~\cite{hartigMultidimensionalIRTModels2009} though factorization directly
within the nonlinear IRT model is possible~\cite{changProbabilisticallyautoencodedHorseshoedisentangledMultidomain2019}.

In this manuscript, we assume that one has designed a test constituted of ordinal response items,
 intended to measure a latent construct in a population. 
 We assume that the items in the test
 have already been pre-partitioned, through domain-specific knowledge, empirical methods,
 or a combination of both.
 Note however that the methods in this manuscript may be used to evaluate
 item partitioning schemes, in case multiple possibilities are being considered.
 Additionally, we assume that one has already administered the test to a  representative sample of individuals within the target population,
 collecting responses that we will refer to as the calibration data. 
 Fundamentally, we assume that one intends to use the data to create an instrument that will generalize to the wider target population.
 At this stage, one has the requisite materials
 needed for model calibration.

\subsection*{Model calibration}

The model in Eq.~\ref{eq:GRM}   can be learned in many ways.
Non-Bayesian approaches typically involve maximum likelihood estimation (MLE)~\cite{lordMaximumLikelihoodEstimation1983,wingerskyInvestigationMethodsReducing1983} or the empirical Bayesian procedure of maximum
marginal likelihood (MML)~\cite{bockIRTEstimationDomain1997},
where given a set of responses $\{x_{pi}\}_{p,i}$ the marginal likelihood
\begin{eqnarray}\label{eq:marginalGRM}
\prod_p \prod_i \Pr( X_{pi} = x_{pi} \vert   \boldsymbol\tau , \boldsymbol\lambda  ) = 
    \prod_p \prod_i \int \Pr( X_{pi} = x_{pi} \vert \theta_p, \boldsymbol\tau , \boldsymbol\lambda  ) \dd\mathcal{N}_{\theta_p}(\hat{\theta}_p,\hat{\sigma}_p)
\end{eqnarray}
is optimized recursively along with estimates of ability,
typically using expectation maximization (EM). 
In Eq.~\ref{eq:marginalGRM}, 
the distribution of ability for a person $p$ is approximated using a Gaussian distribution
centered at $\hat{\theta}_p$ of variance $\hat{\sigma}^2_p$.
Hence, model calibration yields estimates
for both item-specific parameters and person-specific parameters.

\subsubsection*{Bayesian calibration}

Fundamentally, the GRM is a high-dimensional nonlinear latent variable model informed using discrete observations. Hence, its fit must be constrained in order to ensure parametric identifiability. The standard methods for constraining its parameters typically lie in scaling of the ability parameters across the calibration sample so that they are unit scale. Beyond this imposition of scaling, most non-Bayesian methods do not use other regularization.

Modern high-dimensional statistical problems have necessitated the development of regularization techniques. In the Frequentist world, these regularization techniques usually involve penalty functions placed on the objects to be inferred, or hierarchical structure built into the problem as in the case of mixed effects linear regression. Using well-designed regularization, one obtains models that perform better at making predictions than those without regularization. This fact has to do with shrinkage and partial pooling properties of regularized models. Shrinkage is a statistical property where estimates of effect sizes (parameters) are shrunk towards zero. Partial pooling is a form of shrinkage where collectively the differences between parameters are shrunk so that  overall group-level means are obeyed. These properties lead to more robust calibration of models in their response to noisy data. At that point, the lines between Frequentist and Bayesian methods are blurred as regularization can be interpreted as prior information.

In Bayesian modeling of IRT problems, prior distributions are placed
on all model parameters~\cite{lordMaximumLikelihoodEstimation1983}. 
\add{Even when using diffuse priors,} these distributions help regularize the overall inference problem~\cite{lordMaximumLikelihoodBayesian1986,kieftenbeldRecoveryGradedResponse2012}.
Bayesian
modeling allows wide latitude in how one wishes to specify the 
IRT model. For the purposes of this manuscript, we consider modeling
under the principle of using weakly informative prior distributions~\cite{lemoineMovingNoninformativePriors2019},
for the purpose of parameter regularization. To put this principle
in concrete terms, we consider the overall probabilistic model
for generating $X_{pid}$, person $p$'s response to item $i$ in
domain $d$, on a test where each item has $J$ possible responses,
\begin{align}
        X_{pid}&\sim\textrm{categorical}\left(p^{(d)}_{pi,1},\ldots, p^{(d)}_{pi,J}\right) \nonumber \\
        p^{(d)}_{pi,j}  &=\Pr\left( X_{pid} = j \vert \{\theta_p^{(d)}\}_d, \{\{\tau_{ij}^{(d)}\}_j\}_d , \{\lambda_i^{(d)}\}_d  \right) \qquad \textrm{(see Eq.~\ref{eq:GRM})} \nonumber\\
        \lambda_i^{(d)} &\sim \mathrm{cauchy}^+\left(0,5\right)  \nonumber\\
        \tau_{i,j}^{(d)} &\sim \mathrm{normal}^+(\mu_\tau^{(d)},\sigma_\tau^{(d)}) \quad \textrm{subject to } \tau_{i,j}^{(d)} > \tau_{i,j-1}^{(d)} \nonumber\\
        \mu_\tau^{(d)} &\sim \mathrm{normal}(0,5) \nonumber\\
        \sigma_\tau^{(d)} &\sim\mathrm{cauchy}^+(0,1) \nonumber \\
        \theta_p^{(d)}  &\sim \mathrm{normal}(0,{\sigma^{(d)}}) \nonumber\\
        \sigma^{(d)}&\sim\mathrm{cauchy}^+(0,1),
\label{eq:bayesian_irt}
\end{align}
where \sout{$\mathcal{N}$ refers to the normal distribution parameterized by mean and standard deviation, and} $\mathrm{normal}^+, \mathrm{cauchy}^+$ refer to the non-negative half--normal and half--cauchy distributions respectively.
The model presented in Eq.~\ref{eq:bayesian_irt}, features weakly-informative priors that are similar to those used in the prior literature~\cite{luoPerformancesLOOWAIC2017,luoUsingStanProgram2018}.
The purpose of weakly informative priors is to regularize the statistical and computational problem of inferring the parameters of the model. \add{Weakly informative priors have been shown in many contexts to reduce type-I and type-M errors in statistical estimation problems}~\cite{lemoineMovingNoninformativePriors2019,gelmanPriorCanOften2017}, \add{including in IRT problems}~\cite{shengInvestigatingWeaklyInformative2017}.

Bayesian inference involves computing the statistics of a hierarchical model.
For IRT models, this inference is not
analytically tractable and the computations are done typically through approximate methods like Markov Chain Monte
Carlo (MCMC) or Automatic Differentiation Variational Inference (ADVI)~\cite{bleiVariationalInferenceReview2017,natesanBayesianPriorChoice2016}.

 The end result of calibration are posterior distributions over the model parameters in the
Bayesian setting, or point-estimates in the non-Bayesian setting. Point estimates can also be obtained from the 
Bayesian model through consideration of an appropriate loss function. For example, the posterior mean 
minimizes $L^2$ loss whereas the posterior median minimizes $L^1$ loss. 
In many applications, such as the one motivating this manuscript,
point-estimate summaries of the model parameters are necessary.

\subsection*{Scoring}

While calibration is performed off line, the model is usually intended for online use in the aim of
determining the ability scores of new applicants. \add{This process is known as \emph{scoring} and pertains to estimation of the model's latent factors (person-specific ability parameters) conditional on learned posteriors for the item parameters}.

In computer adaptive testing, the calibrated IRT model also guides the presentation of items. 
Classical methods for item selection have used the local Fisher information matrix, conditional on estimated score~\cite{wainerComputerizedAdaptiveTesting2000}. More-contemporary approaches also take score uncertainty into account~\cite{veldkampRobustComputerizedAdaptive2019,changGlobalInformationApproach1996,chengEstimationTraitLevel2000}.
Regardless of item selection approach, a method for scoring is necessary.

In scoring, full Bayesian treatment of the calibrated model parameters is often infeasible
and optimization of the likelihood conditional on point-estimates of the item parameters is a common procedure. 
As opposed to calibration, in this step the item-specific
parameters are assumed known and fixed. 
It is also computationally expensive to propagate posterior distributions of the item parameters so their uncertainty is ignored.
Using their point estimates, the likelihood is maximized relative
to the ability estimate of a new person, given his or her pattern of item responses~\cite{hartigMultidimensionalIRTModels2009}.

To make meaningful comparisons between the ability of people, one must quantify the precision of the ability estimate. 
Given the fitted scores, it is common to perform an asymptotic approximation of the standard error using the Fisher information matrix $I(\hat\theta_p)$. This approximation is the Cramer-Rao lower bound,
\begin{equation}
\textrm{Var}(\hat\theta_p) \geq \frac{1}{I(\theta_p)},  \label{eq:cramer_rao}
\end{equation}
an asymptotic bound for the variance for unbiased estimators based on applying Laplace's method on the 
posterior density.
It is handy to interpret the ability estimate as if it were Gaussian, using the implied variance estimate of Eq.~\ref{eq:cramer_rao}, and the mapping
\begin{equation}
    (\hat{\theta}_p, \hat{s}_p) \mapsto \mathcal{N}_{\theta_p}(\hat{\theta}_p, \hat{s}_p),
\end{equation}
where $\hat{s}_p = 1/\sqrt{I(\hat\theta_p)}.$

\subsubsection*{Weighted Likelihood Estimation (WLE) scoring}
The commonly-used weighted likelihood estimator~\cite{warmWeightedLikelihoodEstimation1989} (WLE)  removes the leading-order asymptotic bias of the maximum likelihood estimator. The asymptotic bias of this particular estimator is $\mathcal{O}(n^{-1})$. 
This estimator is often used in conjunction with the variance estimate of Eq.~\ref{eq:cramer_rao}.

\subsubsection*{Marginal Maximum  Likelihood (MML) scoring}
We also consider an estimator found by maximizing the marginal maximum  likelihood of Eq.~\ref{eq:marginalGRM} with respect to the score directly, hereby referred to as the MML scoring estimator. 
For this estimator, we optimize Eq.~\ref{eq:marginalGRM} with respect to the score ($\hat{\theta}_p$)
while using Eq.~\ref{eq:cramer_rao} to impose the score variance. 
This estimator resembles a variational Bayesian estimator under interchange
of expectations and logarithms within the objective.

\subsubsection*{Expected A Posteriori (EAP) scoring with a compactly-supported prior}

Finally, we consider expected a-posterior (EAP) estimation, using a fully Bayesian procedure where we regularize score inference using a prior distribution. In particular,
we use an explicitly-truncated normal distribution to restrict score estimation within a compact interval surrounding zero. We believe this restriction to be well-motivated when
one realizes that in scoring an individual, the model is interpolating that individual into the pre-calibrated model, finding a placement for that individual relative to the people in the original calibration sample.

In calibration, scores for a representative
sample of a population are inferred. These scores follow some distribution, however,
the usual assumption is that the scores are approximately normal.
Since the scale of the distribution is arbitrary, let us assume without generality 
that the population follows a unit normal distribution on a given trait. 

The tails for a normal distribution fall off rapidly. One should expect to observe someone with scores more extreme than four standard deviations once out of approximately sixteen thousand times. That ratio becomes one in 1.7 million outside of five standard deviations. Hence, for an IRT instrument calibrated using 1.7 million respondents, one would not expect to see scores more extreme than $\pm 5$, when placed on unit scale.

\subsection*{Predictive model evaluation}

This manuscript evaluates model calibration and scoring methods for the GRM based on the predictive power
of each model's corresponding Gaussian approximations of ability (Eq.~\ref{eq:normal}).
In generality, one measures the predictive power of a model by approximating an appropriate risk function as computed by the model on new data. The typical ways of doing this task are cross-validation, and approximate cross-validation through the computation of information criterion. 

Cross-validation involves the separation of datasets into training and testing sets, where the testing set is left out and the model is fit using the training set. The testing set is then used to test the model for predictive accuracy. Information criteria in the cases of the Akaike information criterion (AIC) and Watanabe-Akaike information criterion (WAIC)~\cite{watanabeAsymptoticEquivalenceBayes2010} are \sout{asymptotic approximations of this procedure} \add{ under certain conditions}~\cite{stoneAsymptoticEquivalenceChoice1977} \add{ asymptotic approximations of forms of cross validation}. \add{Regardless, the objective of each of these approaches is to approximate the log likelihood of the model for future data that is not available at training.}

A limitation of each of the information criterion is that they can only be used when making comparisons between models based on the same data. For this reason, they cannot be used for looking at inclusion or exclusion of items since two models with different items use different data. Furthermore, the AIC and WAIC are from different statistical paradigms. The WAIC~\cite{watanabeWidelyApplicableBayesian2013, gelmanUnderstandingPredictiveInformation2014, changPredictiveBayesianSelection2019} is a Bayesian variant of the AIC, scaled to model deviance like the AIC. However, it operates under the assumption that one uses the full posterior of a Bayesian model in making predictions. In the scoring step at test administration, computational trade-offs must be made. While calibration is performed off line, the model is intended for on-line use in determining the ability scores of new applicants. In this stage of computation, Bayes is often infeasible, and optimization of the likelihood is a common procedure. As opposed to calibration, in this step the item-specific parameters are assumed known and fixed. It is also computationally expensive to propagate posterior distributions of the item parameters, so their uncertainty is ignored.

Let $\Omega = \cup_{k=1}^K \Omega^{(k)}$, where $\Omega^{(j)}\cap\Omega^{(k)}=\{\}$ for $j\neq k$,
represent a partition of the $P$ people that responded to items for calibration. 
Leaving out one of the partitions $\Omega^{(k)}$ at a
time, one calibrates (fits) $K$ sets of model parameters. 
The outputs of these calibrations are discrimination parameters $\lambda^{\vert \Omega\setminus\Omega^{(k)}}$ and 
item threshold parameters $\tau^{\vert \Omega\setminus\Omega^{(k)}}$. 
Each calibration, applied to its corresponding left-out data,
yields a set of ability estimates and estimates for the standard deviation of these ability estimates
$
\{
    (
        \hat\theta_p^{\vert \Omega\setminus\Omega^{(k)}}
        , \hat{s}_p^{\vert \Omega\setminus\Omega^{(k)}}
    )
\}_{p\in\Omega^{(k)}}.
$
These estimates self-consistently model the likelihood of the item responses of the left-out people by providing Gaussian approximations
for their abilities $\theta_p$ through the mapping 
\begin{equation}
     (
     \hat\theta_p^{\vert \Omega\setminus\Omega^{(k)}},
    \hat{s}_p^{\vert \Omega\setminus\Omega^{(k)}}
 ) \mapsto \mathrm{normal}_{\theta_p} \left(
     \hat\theta_p^{\vert \Omega\setminus\Omega^{(k)}},
    \hat{s}_p^{\vert \Omega\setminus\Omega^{(k)}}
    \right) ,\label{eq:normal}
\end{equation}
where $\mathcal{N}_\theta(\mu,\sigma^2)$ is a Gaussian measure with mean $\mu$ and variance $\sigma^2$.
This mapping is crucial since it allows one to evaluate comparisons by evaluating quantities such as
$\Pr( \theta_p - \theta_q > 0),$ while respecting uncertainty in ability estimates.

We wish to evaluate a model based on its predictive ability to forecast item response patterns in a way
that is self consistent with such comparisons.
To do so, we use the inferred approximations over the ability distributions and the GRM likelihood to formulate a
prediction risk for the given model. 

 For risk, we consider an approximation of the information loss for a given model $M$, which is expressed by
the deviance-scaled \sout{criterion} \add{measure}
\begin{align}
D[M] &=-2 \sum_k \sum_{p\in\Omega^{(k)}}\log \prod_{i=1}^J \Pr( X_{pi} = x_{pi} \vert  \tau^{\vert \Omega\setminus\Omega^{(k)}}, \lambda^{\vert \Omega\setminus\Omega^{(k)}} )  \nonumber\\
&\approx  -2 \sum_k \sum_{p\in\Omega^{(k)}} \sum_{i=1}^J \log\Bigg[ J_{\hat\theta_p^{\vert \Omega\setminus\Omega^{(k)}},\hat{s}_p^{\vert \Omega\setminus\Omega^{(k)}}}(\lambda^{\vert \Omega\setminus\Omega^{(k)}}_i,\tau^{\vert \Omega\setminus\Omega^{(k)}}_{i,x_{pi}}) \nonumber\\
&\qquad-  J_{\hat\theta_p^{\vert \Omega\setminus\Omega^{(k)}},\hat{s}_p^{\vert \Omega\setminus\Omega^{(k)}}}(\lambda^{\vert \Omega\setminus\Omega^{(k)}}_i,\tau^{\vert \Omega\setminus\Omega^{(k)}}_{i,x_{pi}+1})  \Bigg].
\label{eq:criterion}
\end{align}
The approximation
\begin{equation}
J_{\theta,\sigma}(\lambda,\tau) = \int  \frac{ \dd\mathcal{N}_\phi(\theta, \sigma)}{1+\exp\left(-\lambda(\phi-\tau)\right)} \approx  \Phi\left( \frac{\pi\lambda(\theta-\tau)/8}{\sqrt{1+(\pi\lambda\sigma/8)^2}}\right),
\label{eq:Japprox}
\end{equation}
where \add{ $\mathcal{N}$ is a Gaussian measure } and $\Phi$ is the cumulative density function for the unit normal distribution, 
is well-known~\cite{maragakisBayesianEstimatesFree2008}.

We use the criterion in Eq.~\ref{eq:criterion} to compare regularized Bayesian calibration of the GRM using the model of
Eq~\ref{eq:bayesian_irt} to calibration performed using  marginal maximum
likelihood (MML), coupled with the following scoring methods: expected a-posteriori (EAP), weighted maximum likelihood (WLE)~\cite{warmWeightedLikelihoodEstimation1989}, and marginal maximum  likelihood (MML).
 The criterion being deviance-scaled has the same interpretation as the AIC, consisting of an estimate of the out-of-sample model likelihood.

\section*{Results}
We performed all analyses in R 3.5, with empirical-Bayesian and non-Bayesian analyses performed using the R package \texttt{mirt}. For Bayesian analyses, we used the Stan probabilistic programming language~\cite{carpenterStanProbabilisticProgramming2017}, interfaced in R using the package \texttt{rstan}. We implemented the custom EAP scoring method in R, as well as all of the computations behind the \sout{model deviance criteria} \add{cross-validation metric}  that we use to compare models. \add{This section mainly shows the results of the comparisons performed on both cross-validation folds and on a control sample. The detailed discussion of the results is provided in the Discussion section}.

\subsection*{Comparison of calibration methods on claimant sample}

Our main objective is to compare the predictive performance of regularized
Bayesian calibration to unregularized MML calibration across the different scoring methods (EAP, WLE, MML).
To this end, we used response data collected from the target subpopulation of claimants and calibrated the WD-FAB
using the Bayesian model of Eq.~\ref{eq:bayesian_irt} and using MML.
To be specific,
we used four-fold cross validation on each domain (BM, CC, CM, FMF, ME, RS, SR, UBF), leaving out one fold at a time and fitting the model on the remaining responses.
From each model, we computed the cross-validation criteria for each left-out fold. The only exception was RS, where we instead used three-fold cross validation.

\begin{figure}[!h]
    \centering
    \includegraphics[width=0.8\textwidth]{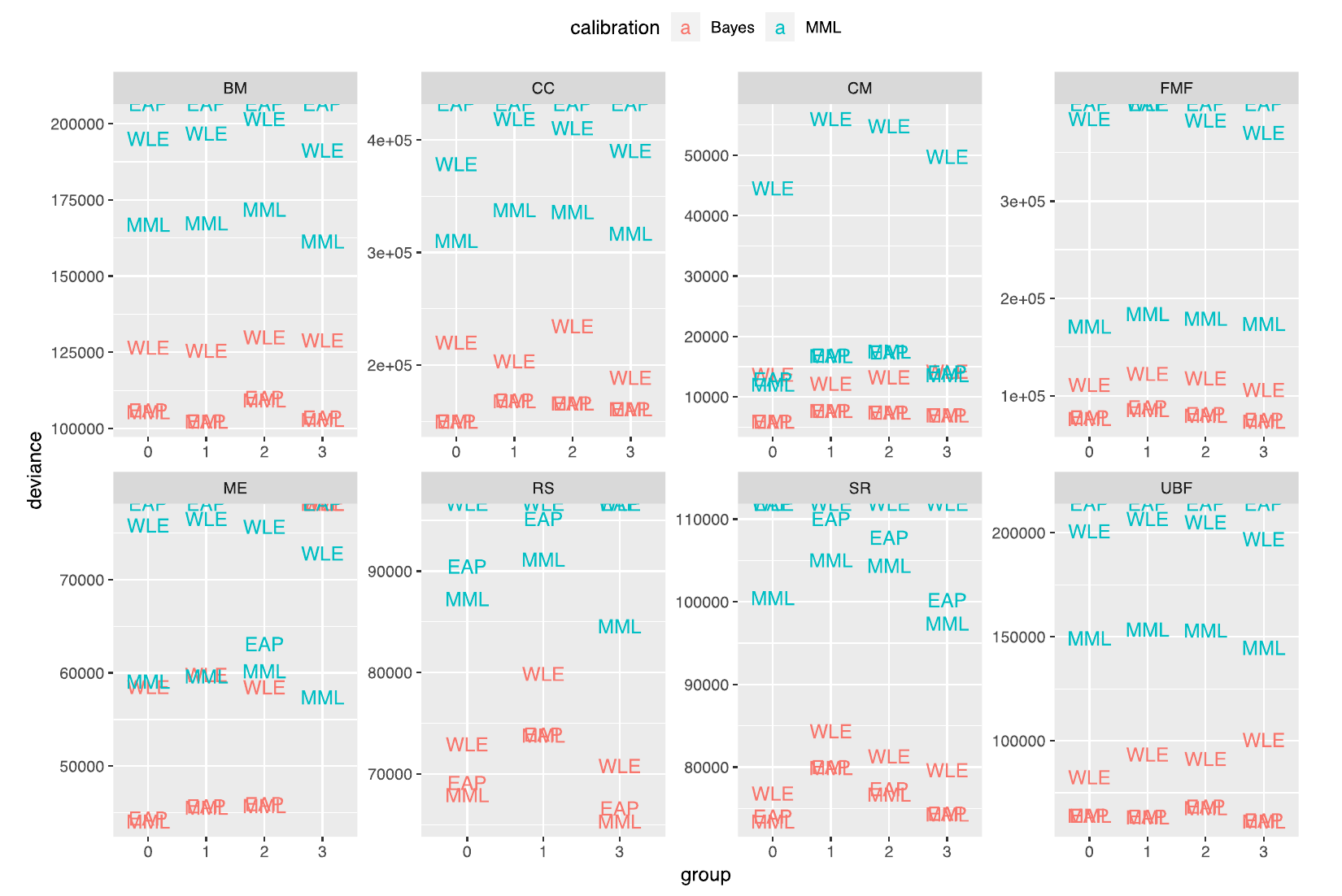}
    \caption{{\bf Four fold cross-validation comparison of calibration of and scoring choices for 8 domains of the WD-FAB (lower is better), performed on claimant sample.} The deviance measure is computed for each left out fold using  Eq.~\ref{eq:criterion}. This measure is scaled in-line with the AIC so lower values are better.}
    \label{fig:fab_deviance}
\end{figure}

\begin{figure}[!h]
    \centering
    \includegraphics[width=0.8\textwidth]{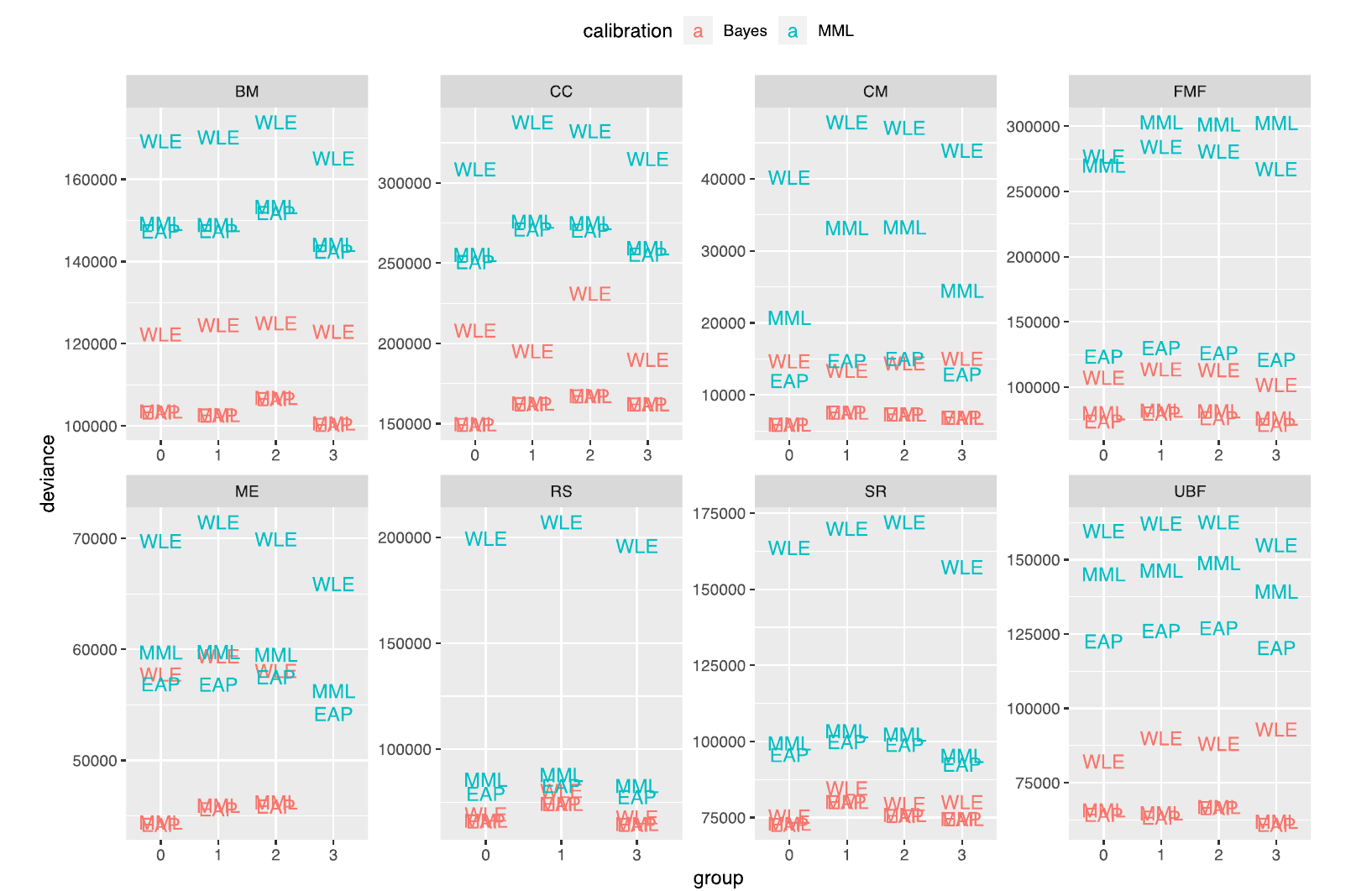}
    \caption{{\bf Point-estimated out-of-sample comparison of calibration and scoring choices on claimant responses for 8 domains of the WD-FAB (lower is better).} The point-estimated deviance measure is computed by taking the variance parameters in Eq.~\ref{eq:criterion} to zero.  This measure is scaled in-line with the AIC so lower values are better.}
    \label{fig:fab_deviance_point}
\end{figure}

First, we computed the metric of Eq.~\ref{eq:criterion} for all pairings of calibration and scoring methods.
In Fig \ref{fig:fab_deviance}, deviance values of Eq.~\ref{eq:criterion} are 
presented for each of the left-out groups. 
The criterion of Eq.~\ref{eq:criterion} takes uncertainty of the estimated scores into account. 
To look at the predictive accuracy of the point-estimate for the score, ignoring uncertainty, we considered the same deviance measure in Eq.~\ref{eq:criterion}, with the variance taken to be zero, regardless of scoring method.
The results of the point-wise criterion are shown in Fig \ref{fig:fab_deviance_point}.

\subsection*{Comparisons on an out-of-sample control population}

\begin{figure}[!h]
    \centering
    \includegraphics[width=0.8\textwidth]{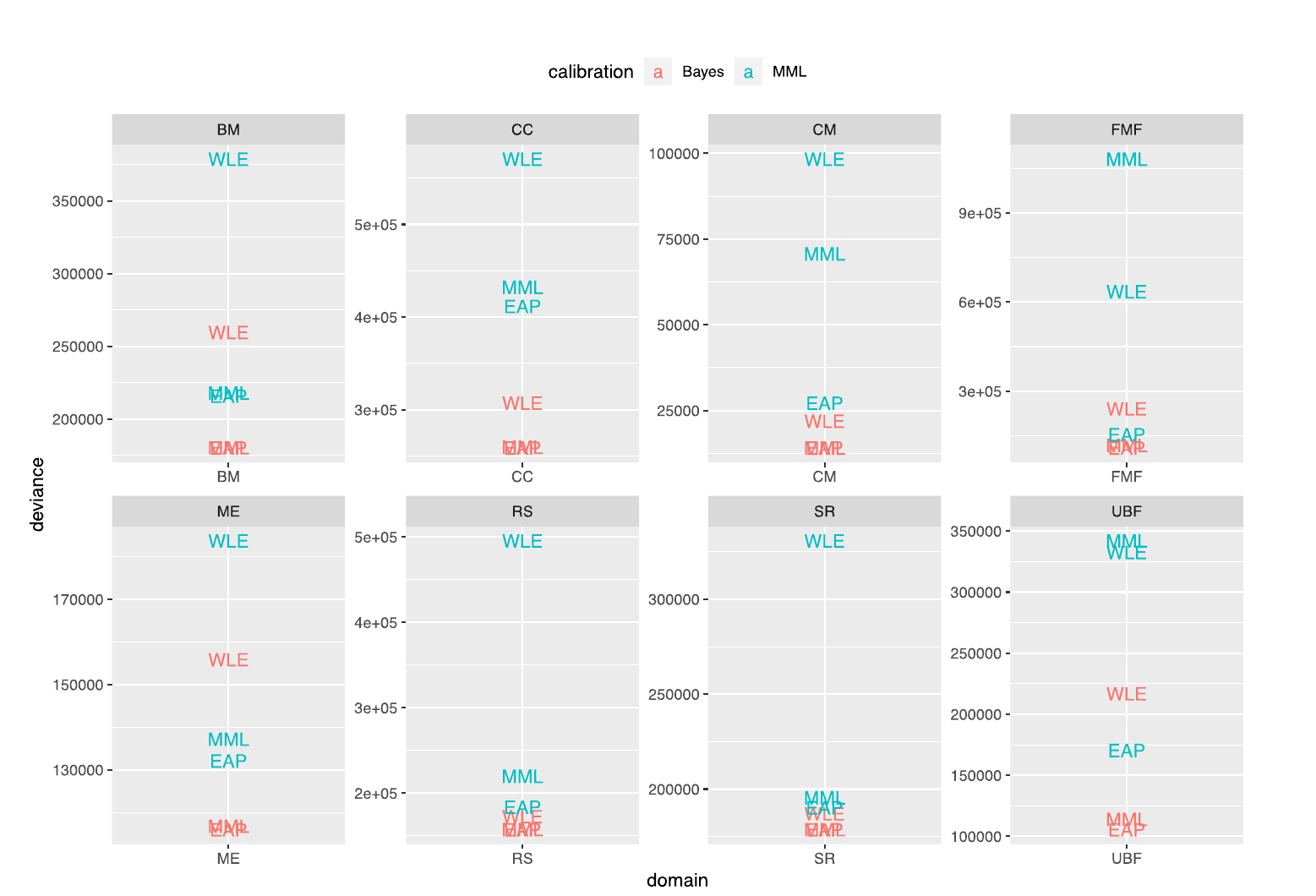}
    \caption{{\bf Out-of-sample out-of-subpopulation comparison of calibration and scoring choices for the 8 domains of the WD-FAB (lower is better).} The deviance measure is given in Eq.~\ref{eq:criterion}. These are computed on a control group that was neither used nor intended for use in calibrating the WD-FAB.}
    \label{fig:fab_deviance_normative}
\end{figure}

For the WD-FAB, the calibration group was taken from a sample of disability claimants.
To put work-related function for these people in context, a separate control
sample of adults was also taken.
This sample was meant to be representative of the population at-large.
We evaluated the different model calibration and scoring methods on the data, using the same measure as in Fig \ref{fig:fab_deviance}, except without leaving out folds during calibration.
Hence, we evaluated how well each model, trained solely on claimant data, predicts responses to the test items for the control population.
The results of this evaluation are shown
in Fig \ref{fig:fab_deviance_normative}.

\subsection*{Comparing scoring methods}

\begin{figure}[!h]
     \centering
    \includegraphics[width=0.8\textwidth]{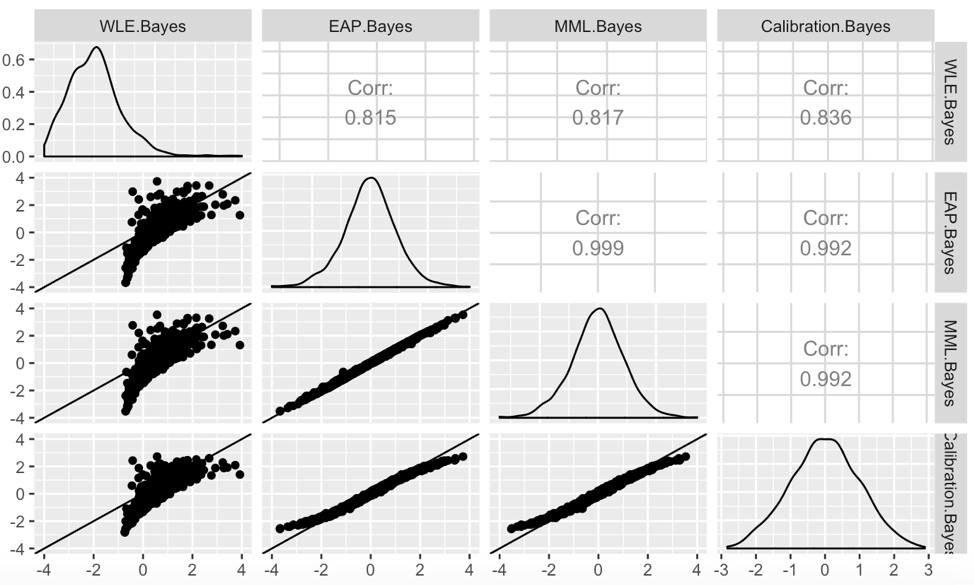}
      \caption{{\bf Pairwise comparison of scoring of the Basic Mobility (BM) scale calibrated using  the full-Bayesian model.} The methods compared are Warm's weighted likelihood estimator (WLE), EAP, and MML. Calibration Bayes refers to the score inferred during item calibration.}
      \label{fig:sim_scoring}
\end{figure}

\begin{figure}[!h]
     \centering
     \includegraphics[width=0.8\textwidth]{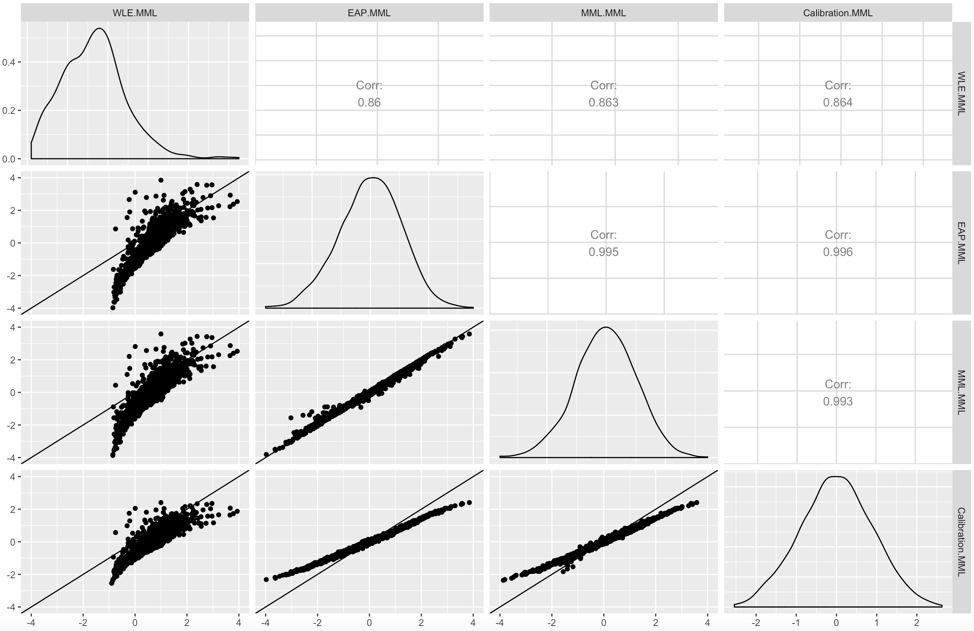}
    \caption{{\bf Pairwise comparison of scoring of the Basic Mobility (BM) scale calibrated using marginal maximum  likelihood (MML).} The methods compared are Warm's weighted likelihood estimator (WLE), EAP, and MML. Calibration MML refers to the score inferred during item calibration.}
     \label{fig:sim_scoring2}
 \end{figure}

Finally, we evaluated each of the  WLE, EAP, and MML scoring methods for consistency with each other and with scores inferred during model calibration.
Recall that calibration also entails inference of abilities -- the item parameters are found self-consistently with these.
Figure \ref{fig:sim_scoring} presents pairwise comparisons of scores obtained using the full Bayesian model, where the label ``Calibration Bayes'' corresponds to posterior means of ability estimates inferred at calibration. Shown are scores for the domain BM.
We performed the same analysis on models calibrated using MML.
These results are shown in Fig \ref{fig:sim_scoring2}. 
Likewise, in this figure, ``Calibration MML'' corresponds to ability estimates for the MML model obtained at calibration.

\section*{Discussion}
In this manuscript we compared full Bayesian inference of IRT models against marginal maximum  likelihood (MML) based empirical Bayesian inference~\cite{harwellItemParameterEstimation1988,ogasawaraMarginalMaximumLikelihood2001}.
Coupled with these these choices for item calibration methodology, we also looked at compactly supported expectation a-posteriori (EAP) scoring compared to weighted likelihood and MML scoring.

Our evaluation metric, targeted at predictive accuracy, is rooted in the concrete and real life application of the assessment of work-related physical and mental function. In particular, the metric is consistent with how scores of such an instrument are typically interpreted - where the error in the ability estimate is interpreted as if it were the standard deviation of a Gaussian distribution. Hence, we defined the metric of Eq.~\ref{eq:criterion} to be consistent with such an interpretation.
In defining the metric, we use the approximation of Eq.~\ref{eq:Japprox}. We note that using probit rather than logistic functions to model the item response functions would render such an approximation unnecessary.
Using this metric we found that choices of calibration and scoring methodologies clearly matter.

\subsection*{Full Bayesian calibration consistently outperforms MML}

In all of our analyses, models calibrated using full Bayesian inference outperformed MML, across all cross-validation splits and all item domains.
This trend is visible in Fig \ref{fig:fab_deviance}, where the Bayesian models have lower estimated model deviance than the MML models, regardless of scoring methodology.
In some applications, one does not care to interpret uncertainty in scoring.
In these situations, one may refer to the analyses of Fig \ref{fig:fab_deviance_point}, where the results are consistent with those of Fig \ref{fig:fab_deviance}. The metric used here more-closely resembles Frequentist cross-validation measures like the AIC that do not account for parameter uncertainty.

Additionally, the Bayesian models transfer better than the MML models.
When evaluating the resulting models on a true out-of-sample set of responses given by the control sample, the same trends held (Fig \ref{fig:fab_deviance_normative}).
While the WD-FAB instrument is meant to be calibrated relative to a claimant subpopulation,
it is not known a-priori whether a given applicant should be a member of this population. Hence, it is important for the instrument to generalize and provide meaningful results for  non-claimants as represented by the control sample.

The superiority of full Bayesian calibration was consistent across all scoring methods, though scoring methods also differed in terms of performance.
The EAP and MML scoring methods exhibited similar performance in all of the cross-validation experiments of Figs.~\ref{fig:fab_deviance} and~\ref{fig:fab_deviance_point}.
WLE, on the other hand, performed consistently poorly when paired with either MML calibration or full Bayesian calibration. 
For many of the domains, pairing the full Bayesian model with WLE scoring was sufficient to remove  the performance advantage of Bayesian calibration over MML calibration.

\subsection*{The WLE approach gives scores inconsistent with item calibration}

Focusing on the poor performance of WLE scoring, relative to the other methods, we compared the scores produced using WLE after calibration versus those of other methods. Figs.~\ref{fig:sim_scoring} and ~\ref{fig:sim_scoring2} provide these comparisons using the full-Bayesian and MML calibrated IRT  models respectively.

The item parameters produced in the calibration of these models are self-consistent with ability estimates produced during calibration.
Hence, it is salient to compare the score obtained with each scoring method against ability estimates produced at calibration.

The WLE scores obtained, conditional on either model, are only weakly-correlated with the scores at calibration.
The WLE is a correction to the first order term in an asymptotic approximation of the signed score bias~\cite{warmWeightedLikelihoodEstimation1989}. 
Our results show that such a correction does more harm than good.
Furthermore, we question the motivations behind attempting to correct this measure of bias, while ignoring other objectives such absolute bias and estimator variance.

\subsection*{EAP regularization for scoring marginally improves predictive performance}

In both Fig \ref{fig:sim_scoring} and Fig \ref{fig:sim_scoring2}, both EAP and MML-based scoring had high correlations ($>0.99$) with scores computed at calibration.
The MML-based scoring method does well - since the objective integrates over an estimate of the score uncertainty, which itself induces shrinkage in the ability estimates. Hence, it is more-regularized than unregularized maximum likelihood or the WLE.

Comparing EAP and MML directly, the EAP method has explicit regularization imposed by the truncated Gaussian prior. 
Hence, while the MML estimator performs some shrinkage, scores computed with EAP tend to be shrunk relative to MML, particularly at the tails.
Imposing compact support on the scoring process guards against facetious extrapolation of the model beyond the range consistent with calibration. By definition, few calibration subjects fall into the extremes. Hence, IRT models are, by construction, less certain in estimating tail behavior within populations.

The shrinkage is most pronounced when performing the comparisons on the MML-calibrated IRT model.
However, looking at the out-of-sample results of Fig \ref{fig:fab_deviance_normative}, we see that EAP generally performs   better than MML in terms of predictive performance as EAP tends to have smaller deviances.
This behavior is expected because the general population has different characteristics than the subpopulation used at calibration.
The regularization in the EAP method helps guard against instabilities induced by these differences. In Fig \ref{fig:sim_scoring}, using the MML-calibrated IRT model, the effects of regularization are clear. EAP shrinks scores but better-preserves relative ordering of scores. In fact, the EAP scores have stronger  correlation than the MML scores are with the scores obtained while calibrating the item parameters using MML.

In this manuscript we compared pairs of calibration and scoring methodologies as applied to assessing predictive ability of the WD-FAB.
In this application, scores and their uncertainty are used to compare respondents.
In-line with the interpretation of the IRT model, we developed the deviance metric of Eq~\ref{eq:criterion}.
We found full-Bayesian item response calibration, coupled with regularized EAP scoring to provide for more-predictive self-consistent model interpretations.

\subsection*{Limitations and extensions}

\add{In this manuscript our goal was to evaluate calibration and scoring of the WD-FAB. For this reason, we did not formulate the cross validation criterion with item selection in mind - or other use cases where two candidate models would have a different set of items. A straightforward method to extend the enclosed methodology to these use cases would be to restrict the sums of Eq.}~\ref{eq:criterion} \add{to shared items. Future work will focus on looking at item selection through a predictive lens.}

\add{For process reasons, we used as a starting point of this work the previously-developed WD-FAB instrument. The development of WD-FAB followed the same procedure as the Patient-Reported Outcomes Measurement Information Systems (PROMIS) program. And all study design and sampling was contingent on the use of this program.

Each unidimensional factor in WD-FAB was developed  from the  exploratory and confirmatory factor analysis. It would be interesting to investigate how WD-FAB can be developed using truly multidimensional IRT models }~\cite{changProbabilisticallyautoencodedHorseshoedisentangledMultidomain2019} \add{which consider the correlation among multiple factors.}

\section*{\sout{Acknowledgements}}
\sout{This work is supported in part by the Intramural Research Program of the National Institutes of Health and the US Social Security Administration.} The opinions expressed in this article are the author's own and do not reflect the view of the National Institutes of Health, the Department of Health and Human Services, US Social Security Administration, or the United States government.

\nolinenumbers

\bibliography{irtvae}

\begin{thebibliography}{10}

\bibitem{changItemResponseTheory2005}
Chang CH, Reeve BB.
\newblock Item {{Response Theory}} and Its {{Applications}} to
  {{Patient}}-{{Reported Outcomes Measurement}};28(3):264--282.
\newblock doi:{10.1177/0163278705278275}.

\bibitem{friesItemResponseTheory2014}
Fries JF, Witter J, Rose M, Cella D, Khanna D, Morgan-DeWitt E.
\newblock Item {{Response Theory}}, {{Computerized Adaptive Testing}}, and
  {{PROMIS}}: {{Assessment}} of {{Physical Function}};41(1):153--158.
\newblock doi:{10.3899/jrheum.130813}.

\bibitem{kingstonFeasibilityUsingItem1982}
Kingston NM, Dorans NJ.
\newblock The {{Feasibility}} of {{Using Item Response Theory}} as a
  {{Psychometric Model}} for the {{Gre Aptitude Test}};1982(1):i--148.
\newblock doi:{10.1002/j.2333-8504.1982.tb01298.x}.

\bibitem{carlsonItemResponseTheory2013}
Carlson JE, von Davier M.
\newblock Item {{Response Theory}};2013(2):i--69.
\newblock doi:{10.1002/j.2333-8504.2013.tb02335.x}.

\bibitem{altemeyerEnemiesFreedomUnderstanding1988}
Altemeyer B.
\newblock Enemies of Freedom: {{Understanding}} Right-Wing Authoritarianism.
\newblock Enemies of Freedom: {{Understanding}} Right-Wing Authoritarianism.
  {Jossey-Bass};.

\bibitem{goldbergDevelopmentMarkersBigFive1992}
Goldberg LR.
\newblock The Development of Markers for the {{Big}}-{{Five}} Factor
  Structure;4(1):26--42.
\newblock doi:{10.1037/1040-3590.4.1.26}.

\bibitem{sullivanAssessmentAlcoholWithdrawal1989}
Sullivan JT, Sykora K, Schneiderman J, Naranjo CA, Sellers EM.
\newblock Assessment of {{Alcohol Withdrawal}}: The Revised Clinical Institute
  Withdrawal Assessment for Alcohol Scale ({{CIWA}}-{{Ar}});84(11):1353--1357.
\newblock doi:{10.1111/j.1360-0443.1989.tb00737.x}.

\bibitem{changProbabilisticallyautoencodedHorseshoedisentangledMultidomain2019}
Chang JC, Vattikuti S, Chow CC. Probabilistically-Autoencoded
  Horseshoe-Disentangled Multidomain Item-Response Theory Models;.
\newblock Available from: \url{http://arxiv.org/abs/1912.02351}.

\bibitem{marinoWorkrelatedMeasuresPhysical2015}
Marino ME, Meterko M, Marfeo EE, McDonough CM, Jette AM, Ni P, et~al.
\newblock Work-Related Measures of Physical and Behavioral Health Function:
  {{Test}}-Retest Reliability;8(4):652--657.
\newblock doi:{10.1016/j.dhjo.2015.04.001}.

\bibitem{jetteWorkDisabilityFunctional2019}
Jette AM, Ni P, Rasch E, Marfeo E, McDonough C, Brandt D, et~al.
\newblock The {{Work Disability Functional Assessment Battery}}
  ({{WD}}-{{FAB}});30(3):561--572.
\newblock doi:{10.1016/j.pmr.2019.03.004}.

\bibitem{marfeoDevelopmentInstrumentMeasure2013}
Marfeo EE, Ni P, Haley SM, Jette AM, Bogusz K, Meterko M, et~al.
\newblock Development of an Instrument to Measure Behavioral Health Function
  for Work Disability: Item Pool Construction and Factor
  Analysis;94(9):1670--1678.

\bibitem{marfeoDevelopmentNewInstrument2016}
Marfeo E, Ni P, Meterko M, Marino M, Peterik K, McDonough C, et~al.
\newblock Development of a {{New Instrument}} to {{Assess Work}}-{{Related
  Function}}: {{Work Disability Functional Assessment Battery}}
  ({{WD}}-{{FAB}});70:7011500012p1--7011500012p1.
\newblock doi:{10.5014/ajot.2016.70S1-RP402B}.

\bibitem{marfeoImprovingAssessmentWork2018}
Marfeo EE, Ni P, McDonough C, Peterik K, Marino M, Meterko M, et~al.
\newblock Improving {{Assessment}} of {{Work Related Mental Health Function
  Using}} the {{Work Disability Functional Assessment Battery}}
  ({{WD}}-{{FAB}});28(1):190--199.

\bibitem{fergusonExploratoryFactorAnalysis1993}
Ferguson E, Cox T.
\newblock Exploratory {{Factor Analysis}}: {{A Users}}’{{Guide}};1(2):84--94.
\newblock doi:{10.1111/j.1468-2389.1993.tb00092.x}.

\bibitem{goretzkoExploratoryFactorAnalysis2019}
Goretzko D, Pham TTH, Bühner M.
\newblock Exploratory Factor Analysis: {{Current}} Use, Methodological
  Developments and Recommendations for Good
  Practice;doi:{10.1007/s12144-019-00300-2}.

\bibitem{williamsExploratoryFactorAnalysis2010}
Williams B, Onsman A, Brown T.
\newblock Exploratory Factor Analysis: {{A}} Five-Step Guide for Novices;8(3).
\newblock doi:{10.33151/ajp.8.3.93}.

\bibitem{howardReviewExploratoryFactor2016}
Howard MC.
\newblock A {{Review}} of {{Exploratory Factor Analysis Decisions}} and
  {{Overview}} of {{Current Practices}}: {{What We Are Doing}} and {{How Can We
  Improve}}?;32(1):51--62.
\newblock doi:{10.1080/10447318.2015.1087664}.

\bibitem{vanprooijenConfirmatoryAnalysisExploratively2001}
van Prooijen JW, van~der Kloot WA.
\newblock Confirmatory {{Analysis}} of {{Exploratively Obtained Factor
  Structures}};61(5):777--792.
\newblock doi:{10.1177/00131640121971518}.

\bibitem{dewaltEvaluationItemCandidates2007}
DeWalt DA, Rothrock N, Yount S, Stone AA.
\newblock Evaluation of {{Item Candidates}}: {{The PROMIS Qualitative Item
  Review}};45:S12--S21.
\newblock doi:{10.1097/01.mlr.0000254567.79743.e2}.

\bibitem{samejimaEstimationLatentAbility1969}
Samejima F.
\newblock Estimation of Latent Ability Using a Response Pattern of Graded
  Scores;34:100--100.

\bibitem{warmWeightedLikelihoodEstimation1989}
Warm TA.
\newblock Weighted Likelihood Estimation of Ability in Item Response
  Theory;54(3):427--450.
\newblock doi:{10.1007/BF02294627}.

\bibitem{kieftenbeldRecoveryGradedResponse2012}
Kieftenbeld V, Natesan P.
\newblock Recovery of {{Graded Response Model Parameters}}: {{A Comparison}} of
  {{Marginal Maximum Likelihood}} and {{Markov Chain Monte Carlo
  Estimation}};36(5):399--419.
\newblock doi:{10.1177/0146621612446170}.

\bibitem{lordMaximumLikelihoodBayesian1986}
Lord FM.
\newblock Maximum {{Likelihood}} and {{Bayesian Parameter Estimation}} in
  {{Item Response Theory}};23(2):157--162.
\newblock doi:{10.1111/j.1745-3984.1986.tb00241.x}.

\bibitem{gelmanPriorCanOften2017}
Gelman A, Simpson D, Betancourt M.
\newblock The {{Prior Can Often Only Be Understood}} in the {{Context}} of the
  {{Likelihood}};19(10):555.
\newblock doi:{10.3390/e19100555}.

\bibitem{shengInvestigatingWeaklyInformative2017}
Sheng Y.
\newblock Investigating a Weakly Informative Prior for Item Scale
  Hyperparameters in Hierarchical {{3PNO IRT}} Models;8:123.
\newblock doi:{10.3389/fpsyg.2017.00123}.

\bibitem{lemoineMovingNoninformativePriors2019}
Lemoine NP.
\newblock Moving beyond Noninformative Priors: Why and How to Choose Weakly
  Informative Priors in {{Bayesian}} Analyses;128(7):912--928.
\newblock doi:{10.1111/oik.05985}.

\bibitem{watanabeAsymptoticEquivalenceBayes2010}
Watanabe S.
\newblock Asymptotic {{Equivalence}} of {{Bayes Cross Validation}} and {{Widely
  Applicable Information Criterion}} in {{Singular Learning
  Theory}};11:3571--3594.

\bibitem{watanabeWidelyApplicableBayesian2013}
Watanabe S.
\newblock A {{Widely Applicable Bayesian Information Criterion}};14:867--897.

\bibitem{luoPerformancesLOOWAIC2017}
Luo Y, Al-Harbi K.
\newblock Performances of {{LOO}} and {{WAIC}} as {{IRT Model Selection
  Methods}};59(2):183.

\bibitem{gelmanUnderstandingPredictiveInformation2014}
Gelman A, Hwang J, Vehtari A.
\newblock Understanding Predictive Information Criteria for {{Bayesian}}
  Models;24(6):997--1016.
\newblock doi:{10.1007/s11222-013-9416-2}.

\bibitem{stoneAsymptoticEquivalenceChoice1977}
Stone M.
\newblock An {{Asymptotic Equivalence}} of {{Choice}} of {{Model}} by
  {{Cross}}-{{Validation}} and {{Akaike}}'s {{Criterion}};39(1):44--47.

\bibitem{PromisePROMIS2010}
The {{Promise}} of {{PROMIS}};59(2):77.
\newblock doi:{10.1097/NNR.0b013e3181d7d1b1}.

\bibitem{hartigMultidimensionalIRTModels2009}
Hartig J, Höhler J.
\newblock Multidimensional {{IRT}} Models for the Assessment of
  Competencies;35(2):57--63.
\newblock doi:{10.1016/j.stueduc.2009.10.002}.

\bibitem{lordMaximumLikelihoodEstimation1983}
Lord FM.
\newblock Maximum Likelihood Estimation of Item Response Parameters When Some
  Responses Are Omitted;48(3):477--482.
\newblock doi:{10.1007/BF02293689}.

\bibitem{wingerskyInvestigationMethodsReducing1983}
Wingersky MS, Lord FM.
\newblock An {{Investigation}} of {{Methods}} for {{Reducing Sampling Error}}
  in {{Certain Irt Procedures}}*;1983(2):i--52.
\newblock doi:{10.1002/j.2330-8516.1983.tb00028.x}.

\bibitem{bockIRTEstimationDomain1997}
Bock RD, Thissen D, Zimowski MF.
\newblock {{IRT Estimation}} of {{Domain Scores}};34(3):197--211.
\newblock doi:{10.1111/j.1745-3984.1997.tb00515.x}.

\bibitem{luoUsingStanProgram2018}
Luo Y, Jiao H.
\newblock Using the {{Stan}} Program for {{Bayesian}} Item Response
  Theory;78(3):384--408.

\bibitem{bleiVariationalInferenceReview2017}
Blei DM, Kucukelbir A, McAuliffe JD.
\newblock Variational {{Inference}}: {{A Review}} for
  {{Statisticians}};112(518):859--877.
\newblock doi:{10.1080/01621459.2017.1285773}.

\bibitem{natesanBayesianPriorChoice2016}
Natesan P, Nandakumar R, Minka T, Rubright JD.
\newblock Bayesian {{Prior Choice}} in {{IRT Estimation Using MCMC}} and
  {{Variational Bayes}};7.
\newblock doi:{10.3389/fpsyg.2016.01422}.

\bibitem{wainerComputerizedAdaptiveTesting2000}
Wainer H, Dorans NJ, Flaugher R, Green BF, Mislevy RJ.
\newblock Computerized {{Adaptive Testing}}: {{A Primer}}.
\newblock {Routledge};.

\bibitem{veldkampRobustComputerizedAdaptive2019}
Veldkamp BP, Verschoor AJ.
\newblock Robust {{Computerized Adaptive Testing}}.
\newblock In: Veldkamp BP, Sluijter C, editors. Theoretical and {{Practical
  Advances}} in {{Computer}}-Based {{Educational Measurement}}. Methodology of
  {{Educational Measurement}} and {{Assessment}}. {Springer International
  Publishing};. p. 291--305.

\bibitem{changGlobalInformationApproach1996}
Chang HH, Ying Z.
\newblock A {{Global Information Approach}} to {{Computerized Adaptive
  Testing}};20(3):213--229.
\newblock doi:{10.1177/014662169602000303}.

\bibitem{chengEstimationTraitLevel2000}
Cheng PE, Liou M.
\newblock Estimation of {{Trait Level}} in {{Computerized Adaptive
  Testing}};24(3):257--265.
\newblock doi:{10.1177/01466210022031723}.

\bibitem{changPredictiveBayesianSelection2019}
Chang JC.
\newblock Predictive {{Bayesian}} Selection of Multistep {{Markov}} Chains,
  Applied to the Detection of the Hot Hand and Other Statistical Dependencies
  in Free Throws;6(3):182174.
\newblock doi:{10.1098/rsos.182174}.

\bibitem{maragakisBayesianEstimatesFree2008}
Maragakis P, Ritort F, Bustamante C, Karplus M, Crooks GE.
\newblock Bayesian Estimates of Free Energies from Nonequilibrium Work Data in
  the Presence of Instrument Noise;129(2):07B609.

\bibitem{carpenterStanProbabilisticProgramming2017}
Carpenter B, Gelman A, Hoffman MD, Lee D, Goodrich B, Betancourt M, et~al.
\newblock Stan : {{A Probabilistic Programming Language}};76(1).
\newblock doi:{10.18637/jss.v076.i01}.

\bibitem{harwellItemParameterEstimation1988}
Harwell MR, Baker FB, Zwarts M.
\newblock Item {{Parameter Estimation Via Marginal Maximum Likelihood}} and an
  {{EM Algorithm}}: {{A Didactic}};13(3):243--271.
\newblock doi:{10.3102/10769986013003243}.

\bibitem{ogasawaraMarginalMaximumLikelihood2001}
Ogasawara H.
\newblock Marginal Maximum Likelihood Estimation of Item Response Theory
  ({{IRT}}) Equating Coefficients for the Common-Examinee Design;43(2):72--82.
\newblock doi:{10.1111/1468-5884.00162}.

\end{thebibliography}

\end{document}